\documentclass[sigconf,pbalance]{acmart}

\copyrightyear{2026}
\acmYear{2026}
\setcopyright{cc}
\setcctype{by-nc-nd}
\acmConference[ICAIF '26]{7th ACM International Conference on AI in Finance}{November 14--17, 2026}{Milan, AA, Italy}
\acmBooktitle{7th ACM International Conference on AI in Finance (ICAIF '26), November 14--17, 2026, Milan, AA, Italy}
\acmDOI{10.1145/3858063.3858900}
\acmISBN{979-8-4007-3087-0/2026/11}

\begin{document}

\title{Neural Estimation of Irreversibility in Real and Simulated
Limit Order Book Paths}

\author{Hyeonjun Yeo}
\authornote{First author.}
\orcid{0009-0005-9368-306X}
\affiliation{%
  \institution{Seoul National University}
  \city{Seoul}
  \country{Republic of Korea}}
\email{hj.yeo30@gmail.com}

\author{Doohwi Cha}
\orcid{0009-0000-2513-6965}
\affiliation{%
  \institution{Mirae Asset Securities}
  \city{Seoul}
  \country{Republic of Korea}}
\email{doohwicha@gmail.com}

\author{Yoontae Hwang}
\authornote{Corresponding author.}
\orcid{0000-0002-5856-1914}
\affiliation{%
  \institution{Pusan National University}
  \city{Busan}
  \country{Republic of Korea}}
\email{yoontae.hwang@pusan.ac.kr}

\renewcommand{\shortauthors}{Yeo et al.}

\begin{abstract}
A path indistinguishable from its own reverse carries no arrow of time, so
time irreversibility is a necessary condition for a directed mechanism to be
readable from trajectories alone. In finance it has been measured only on
scalar series such as returns and volatility. We define the irreversibility
rate of a coarse-grained limit order book path as the forward-to-reverse
log-density ratio per event, and estimate it from three weeks of Nasdaq
market-by-order data with an order-invariant MLP and a temporal convolutional
network that receive identical blocks. For NVDA, the order-invariant and
order-sensitive estimates are 0.036 and 0.054 nats per event, respectively.
Their difference, the ordering increment, is 0.018. A surrogate preserving the
coarse one-step transition table recovers 30 percent less irreversibility than
the observed paths. The increment is positive
for all twelve symbols on every held-out daytime date, and for NVDA in each
session window. Across the twelve symbols, its share of the ordered rate
increases roughly logarithmically with the spread in ticks. The increment
changes more than tenfold with the coarse-graining scale. Neither a linear score
nor any single feature group reproduces the full fitted rate.
A queue-reactive simulator calibrated on the same days
matches the market's ordered rate, 0.052 against 0.054, yet its ordering
increment is 0.0006 against 0.018, building temporal asymmetry out of block
composition instead of block order. We propose the ordering and scale profile
of path irreversibility as a benchmark for order book simulators.
Code and audit materials are available at
\href{https://github.com/hjyeo30/lob-irreversibility}%
{\nolinkurl{https://github.com/hjyeo30/lob-irreversibility}}.

\end{abstract}




\maketitle

\section{Introduction}
\label{sec:intro}

% Paper-level overview figure. It is declared here so that it floats to the top
% of the first page that can carry a full-width float, and it also carries the
% reversal-and-estimator schematic that Section 4.1 refers back to.
\begin{figure*}[t]
\centering
\includegraphics[width=\textwidth]{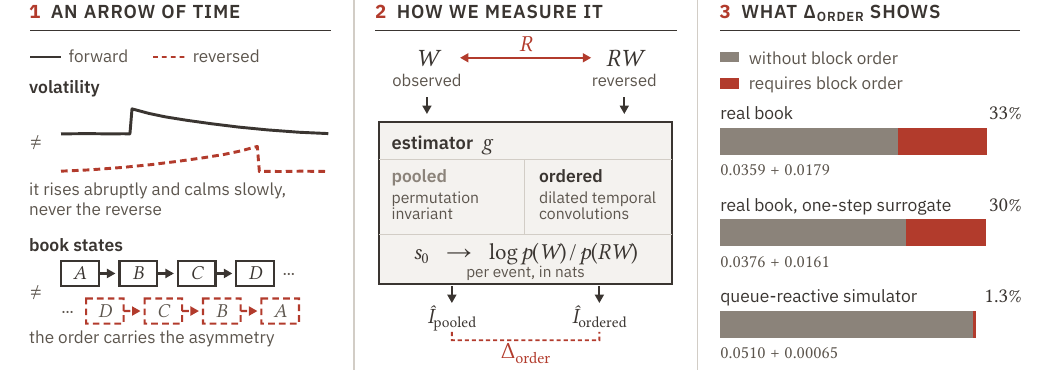}
\Description{Three panels connect time reversal to neural estimation and rate comparisons. At left, forward and reversed traces illustrate volatility and book states. In the middle, pooled and ordered networks estimate forward-to-reverse scores; the networks are trained separately. At right, the real-book and simulator bars combine the pooled rate in gray and the ordered-minus-pooled increment in red. The middle bar instead shows the surrogate rate in gray and the reduction from the observed ordered rate in red. The red shares are 33 percent, 30 percent, and 1.3 percent, respectively.}
\caption{Overview of the measurement, for NVDA daytime at $B=10$ and $K=6$. In
\textbf{(1)}, a forward path and its reversal are drawn for a volatility trace
and for a sequence of book states. In \textbf{(2)}, a path $W$ and its physical
reversal $RW$ are scored by pooled and ordered networks with the same block
inputs and encoder design, trained separately. In \textbf{(3)}, the real-book
and simulator bars show the pooled rate in gray and the ordering increment,
$\Delta_{\mathrm{order}} = \widehat I_{\mathrm{ordered}} - \widehat
I_{\mathrm{pooled}}$, in red. In the middle row, gray denotes the surrogate rate
and red the reduction from the observed ordered rate. All rates are in nats per
event.}
\label{fig:overview}
\end{figure*}

The limit order book is where a market is assembled. Every quote, cancellation
and trade enters it as a message, and liquidity supply, order flow and price
formation meet there on timescales of milliseconds. We ask whether the recorded
path is statistically distinguishable from the path obtained by running the same
book episode in reverse. Time irreversibility measures this distinction from the
path alone.

Irreversibility was developed in statistical mechanics to separate directed
dynamics from equilibrium fluctuation. For a stationary scalar process the
autocovariance satisfies $C(\tau)=C(-\tau)$, so the variance, the power spectrum
and every second-order statistic are unchanged when the series is played
backwards~\cite{weiss1975time}. What distinguishes forward from reverse time is
the ordering of joint fluctuations. An asymmetry as familiar as volatility rising
abruptly and decaying slowly is therefore invisible to those statistics, and
irreversibility is a property of a path distribution rather than of a marginal
one, measured by how far that distribution can be told apart from its
reverse~\cite{parrondo2009entropy}.

In markets it has been examined through volatility coupling across time
scales~\cite{zumbach2009time}, visibility-graph asymmetries in price
series~\cite{lacasa2012time,flanagan2016irreversibility}, classi\-fier-based
estimates on reversed return windows \cite{vodret2024functional}, and multiscale
lead--lag asymmetries in trader-resolved activity~\cite{cordi2021nanostructure}.
These applications read the asymmetry from a scalar series, however, and none of them
constructs a physical reversal of the joint state-and-flow path that the book
records at the level of individual messages. The literature on the book itself
asks different questions. Order-flow imbalance explains a large share of
contemporaneous short-interval price variation~\cite{cont2014price}, queueing and
queue-reactive simulators reproduce important aspects of price and liquidity
dynamics~\cite{cont2013price,huang2015simulating}, and deep sequence models
extract further predictive structure~\cite{zhang2019deeplob}. All of these ask
how the current book predicts a future outcome, or whether a simulator matches
selected market statistics. How much the joint book path reveals about its own
direction of time, and over how many events, has not, to our knowledge, been
measured.

Simulated books carry the same gap. Queue-re\-active and\linebreak[4] learned generative
simulators are calibrated so that their messages, spreads, depths and impact
curves match those of a real venue~\cite{cont2023lobgan}, they now run at a scale
that makes large-scale experimentation
routine~\cite{frey2023jaxlob}, and current benchmarks score them by the distance
between generated and real data~\cite{nagy2025lobbench}. No comparison of that
kind asks whether a generated path could be told apart from its own reverse, so a
simulator may reproduce every marginal that is checked and still move the book
through its states in a different order. For simulator evaluation, this raises
the question of whether a model reproduces the asymmetry associated with
temporal ordering. A simulator may reach a similar overall irreversibility
rate while recovering most of it without access to block order. Comparing the
order-invariant and order-sensitive estimates on real and generated paths makes
this difference measurable.

We therefore construct coarse-grained book paths whose physical reversal changes
the sign of directional variables and exchanges liquidity depletion with
appearance (Fig.~\ref{fig:overview}). We estimate the for\-ward-to-reverse density ratio with an
antisymmetric neural score, following density-ratio estimation by
classification~\cite{sugiyama2012density,gutmann2010noise} and the pathwise
probability-ratio view of irreversibility in stochastic
thermodynamics~\cite{seifert2012stochastic,kim2020learning}. Because an
irreversible path need not owe its asymmetry to temporal order, we compare an
order-invariant neural estimator with an order-sensitive temporal network. Both
receive the same block inputs. We call the difference between their estimated
irreversibility rates the \emph{ordering increment}.

The comparison also raises a question across books: where does order matter
more? We examine whether the increment's share of the ordered rate varies with
the spread in ticks, which describes how tightly the tick constrains the book.

The size of the increment is not fixed by the event horizon, because
coarse-graining discards the within-block ordering, which carries part of the
asymmetry, so the same nominal horizon read at a coarser resolution is not the
same measurement~\cite{esposito2012stochastic}. An irreversibility rate belongs
to a representation and an observation scale rather than to a market. We
summarize our contributions as follows.

\begin{itemize}
\item \textbf{A path-level measure of order book irreversibility.} We define a
physical reversal of the joint state-and-flow path and estimate its
for\-ward-to-reverse log-density ratio with an antisymmetric network. The ordering
increment is positive for all twelve symbols on every
held-out daytime date, and for NVDA in each session window.
Across these symbols, its share of the ordered rate increases
roughly logarithmically with the spread in ticks.

\item \textbf{Comparisons with one-step references and individual feature groups.}
Two references that retain one-step information fall short of the rate learned
from the whole path, and no single feature group reproduces the full fitted rate.

\item \textbf{A within-path benchmark for order book simulators.} A
calibrated queue-reactive simulator reaches the market's ordered rate but almost
none of its ordering increment. We propose reporting how the increment responds
to block order and to coarse-graining scale, the ordering and scale profile, as a
benchmark dimension that generated-to-real comparisons leave untested.
\end{itemize}

\section{Background and Related Work}

This measurement sits between the study of temporal asymmetry in financial series
and the modelling and evaluation of order book dynamics.

\subsection{Time irreversibility in financial dynamics}
\label{sec:tirf}

Zumbach~\cite{zumbach2009time} gives the modern reference point, comparing the
correlation of coarse past vola\-tility with fine future vola\-tility\linebreak[4] against its
mirror image. The stochastic-vola\-tility constructions evaluated there do not
generate the observed gap, and in Markov jump systems an asymmetry of that kind
places a lower bound on the cycle affinity required to sustain
it~\cite{ohga2023thermodynamic}. Later work replaces the chosen statistic
with a general one. Visibility graphs map a series to a network whose in-degree
and out-degree distributions differ under
irreversibility~\cite{lacasa2012time,flanagan2016irreversibility}. Vodret et
al.~\cite{vodret2024functional} estimate a Kullback--Leibler rate by training a
gradient boosting classifier to tell a window from its reverse, and Cordi et
al.~\cite{cordi2021nanostructure} document multiscale lead--lag asymmetries in
trader-resolved activity. Fitting a generative model to the forward sample is
not a direct measure of temporal direction. Univariate and symmetric Hawkes
processes, for instance, assign almost the same likelihood to a sequence and to
its reverse~\cite{cordi2017testing}. From this line of work we keep the
classification idea, which fits no generative model, and move the object of
study to the multivariate book path, where reversal preserves the values of
some features, changes the signs of others and exchanges the roles of paired events.

Separating a trajectory from its reverse is a density-ratio problem, and
training a discriminator is a standard route to such
ratios~\cite{sugiyama2012density,gutmann2010noise}. In stochastic
thermodynamics, path log-ratios under conjugate dynamics quantify entropy
production when local detailed balance holds~\cite{seifert2012stochastic}.
Rold\'an and Parrondo~\cite{roldan2010estimating} estimate dissipation from
stationary trajectory statistics, while Kim et al.~\cite{kim2020learning}
introduce a neural antisymmetric estimator. We take the antisymmetry construction
from this literature and use a logistic density-ratio objective. Where the
physical ratio acquires an energetic meaning through local detailed balance,
ours measures how far the observed market path can be told apart from its
reverse.

\subsection{Modelling and evaluating order book dynamics}

The event-level view of the book is well
developed~\cite{gould2013limit,bouchaud2018trades}. Order-flow imbalance
explains a large share of contemporaneous price
change~\cite{cont2014price,eisler2012price}, and decomposing that flow by event
type sharpens the relation~\cite{sitaru2023orderflow}. Queueing and
queue-reactive simulators reproduce price and liquidity dynamics from
state-dependent intensities~\cite{cont2013price,huang2015simulating}, now also
with learned components~\cite{bodor2025mdqr}. Market-by-order data has also been
used to separate the participants behind the flow, by clustering individual
orders into directional, opportunistic and market-making
groups~\cite{zhang2025clusterlob}. These studies explain or predict price
responses from the state and recent history of the book, or attribute flow to its
sources, whereas irreversibility concerns the joint path and its temporal
direction. We borrow from it the transaction-time sampling
idea~\cite{ane2000order} and the dilated residual temporal
model~\cite{bai2018empirical} applied to order book
sequences~\cite{zhang2019deeplob}.

Evaluating simulated books has become a subfield of its own, as simulators have
grown fast enough to be run at scale~\cite{frey2023jaxlob}. Generative simulators
are judged by whether they reproduce stylized facts and impact
curves~\cite{cont2023lobgan}, benchmarks score generated message data against
real data~\cite{nagy2025lobbench}, and the gap between the two is now an object
of study in itself~\cite{noble2026reality}. They provide the
generated--real comparisons that the within-path measurement of
Section~\ref{sec:benchmark} complements.

\section{Data and Path Representation}

Measuring the irreversibility of a book path requires first deciding what counts
as a path. Any grouping of messages fixes both the reversal operator and the
scale at which asymmetry can appear, and Section~\ref{sec:scale} returns to that
scale dependence as an empirical result.

\subsection{Market data and sample}

We use Nasdaq TotalView-ITCH market-by-order data obtained through Databento
(``XNAS.ITCH'', MBO schema) for the fifteen trading days from 2026-07-02 to
2026-07-23, re\-constructing the\linebreak[4] book by chronological
replay of add, cancel, replace and displayed execution messages. One exchange
action can be published as several messages, as when a marketable order consumes
several resting orders in turn, so we count \emph{source events}, the completed
book updates that the replay resolves, rather than raw messages.

The primary symbol is NVDA and the primary session window is 10:00 to 15:30 New
York time, which we call daytime. The cross-section adds eleven other symbols,
named in Fig.~\ref{fig:cross} and selected to span activity, tick constraint and
queue depth over more than two orders of magnitude in daily event count.

\subsection{Coarse-grained event paths}

Message labels do not define a one-to-one physical reversal. Cancellation and
execution both take a best queue from $Q$ to $Q-v$, whereas their reverse is
liquidity appearing at that queue, whose original cause cannot be recovered.
Grouping messages resolves this ambiguity, since over a block the flow is
summarized by the liquidity added and the liquidity removed and reversal exchanges
the two. A block also records the gross change as well as the net one, because
additions and removals of the same size inside one block leave the boundary states
identical even though the block was not quiet.
We therefore sample the book on an event clock, so a block is an event bar of $B$
source events~\cite{ane2000order}. It targets $B$ of them and closes at the first
boundary where the current exchange action has completed and the book is valid
and uncrossed, which makes the realized count
\begin{equation}
N_t = B + O_t,
\end{equation}
where the completion overshoot $O_t \ge 0$ is recorded as an observed feature.
The overshoot is small, and for NVDA at $B=10$ a nominal horizon of sixty updates
covers about 61 realized events. Blocks never cross a date, a
session-window boundary or a book reset published by the exchange.

Block $t$ is encoded as $z_t=[S_{t-1},U_t,S_t]$, a boundary vector $S$ at each
end and an interval vector $U$ between them. Each block is therefore described by
37 features in four groups, together with an \emph{activity clock} recording
elapsed time, realized event count and completion overshoot.
Book state is the boundary itself, seven values holding spread,
best-quote and five-level depth on each side, and depth imbalance at one and five
levels. Both ends of a block are recorded, so this group contributes fourteen
features. Mechanical flow is the signed net depth change and the gross
depth variation for each side at the best quote and over five levels, eight
features. Liquidity events is the best-quote depletion and appearance
counts on each side with their sum and difference, six features. Price
is the mid log return, the mid change, the bid and ask quote changes, and the
absolute mid and spread variation, six features.
Order identifiers, absolute timestamps, raw
price levels and message cause labels stay out of the network input, since an order
ages monotonically and separates a path from its reverse on its own.

A path of length $K$ is
\begin{equation}
W=(S_0,U_1,S_1,\ldots,U_K,S_K),
\end{equation}
built inside one symbol, date, session window and contiguous segment, with
nominal event horizon $H=BK$. Path start points lie on a fixed stride of three
blocks, so for $K>3$ neighboring paths overlap and held-out dates rather than
paths are the independent repeats. Features are put on a common scale by a
signed logarithmic transform, with per-feature medians and robust scales estimated
on the nine development days of Section~\ref{sec:design} only and shared by forward
and reverse paths, all block positions and all feature groups.

\section{Measuring Irreversibility of Order Book Paths}

Turning a path into a number requires a reversal operator that is faithful to
what each feature means, and an estimator whose output can respond only to what
that operator changes.

\subsection{Physical reversal and antisymmetric estimation}

Reversing a book path means asking what the same episode would look like if the
exchange had run backwards. We write $R$ for the operator that performs this
reversal, and it has to respect what each feature
means (Fig.~\ref{fig:overview}(2)). State levels such as spread, depth and
imbalance keep their values and change position. Interval features that keep
their values under reversal are called \emph{even}, including gross depth
variation, absolute price variation and elapsed time. Features that change sign
are called \emph{odd}, so a mid-price
change of $+1$ tick forward appears as $-1$ tick reversed, and the same holds for
net depth change and bid and ask quote change. Paired event roles exchange,
because liquidity that vanished from the best quote in forward time appears there
in reverse time. Applying $R$ twice returns the path unchanged.

We train a network to separate an observed path from its reverse and read its
calibrated log-odds as an estimate of the log-ratio of the two path densities.
Let $g$ be the raw output of a network on a path. The score we use is
\begin{equation}
s_0(W)=g(W)-g(RW),
\label{eq:antisym}
\end{equation}
which gives $s_0(RW)=-s_0(W)$ for any $g$, without any constraint on the
architecture. Any component of $g$ that is even in the direction of time cancels
in Eq.~\eqref{eq:antisym}, so the score can only respond to what changes under
reversal. Taking the forward path as the positive orientation, we minimize the
paired logistic loss
\begin{equation}
\mathcal{L}=\frac{1}{N}\sum_{i=1}^{N}\log\!\left(1+e^{-s_0(W_i)}\right),
\end{equation}
the standard density-ratio objective~\cite{sugiyama2012density}. The reverse of
each path receives its negative label through the antisymmetry, so reverse rows
are never duplicated into the sample. At the population optimum, balanced
logistic classification yields the
forward-to-reverse log-density ratio $\log[p(W)/p(RW)]$, whose forward
expectation is the Kullback--Leibler divergence between the path distribution and
its reverse. Because finite-network logits can be miscalibrated, on calibration
days we fit one nonnegative scalar, $s(W)=\alpha s_0(W)$, by minimizing the same
loss with days weighted equally. This is temperature
scaling~\cite{guo2017calibration} and leaves the learned representation
untouched.
To limit the influence of extreme logits in sparsely supported regions, we
define
\[
s_c(W)=\operatorname{clip}(s(W),-\log 99,\log 99),
\]
which holds the
implied forward probability inside $[0.01,0.99]$. Primary com\-parisons use $s_c$.
We also report the unclipped $s$ for the spread sensitivity comparison of
Section~\ref{sec:spread} and the length profile of Section~\ref{sec:scale},
where the bound becomes active on most long paths.
Because the score is a log-odds, its natural unit is the nat. Let $\mathcal D$
be the test dates and $\mathcal W_d$ the paths on date $d$. For a reported score
$q\in\{s,s_c\}$ the date-equal event-normalized rate is
\begin{equation}
\widehat I_{\mathrm{event}}(q)
  =\frac{1}{|\mathcal D|}\sum_{d\in\mathcal D}
    \frac{\sum_{W\in\mathcal W_d}q(W)}
         {\sum_{W\in\mathcal W_d}N(W)},
\end{equation}
where $N(W)=\sum_t N_t$ is the realized event count of the path. This reports
paths of different lengths and resolutions in a common unit, and unless stated
otherwise \emph{irreversibility rate} refers to
$\widehat I_{\mathrm{event}}(s_c)$.

\subsection{Order-invariant and order-sensitive networks}

An irreversible path distribution need not depend on the order of its parts. If
the asymmetry rested entirely on the multiset of block contents, a summary blind
to sequence would capture it.
Both architectures use the same block-encoder design, a two-layer perceptron that
maps the 37 features of a block to a 32-dimensional embedding. Each network trains
its own copy of the encoder, whose weights are shared across all block positions
within that network.

The order-invariant member of the pair, \emph{pooled}, concatenates
the mean and the maximum of the $K$ block
embeddings and passes them through a small readout. Pooling makes it invariant to
any permutation of the blocks~\cite{zaheer2017deep}, so it expresses which kinds
of block content a path contained and in what mixture, and nothing about
sequence. The order-sensitive member, which we call \emph{ordered}, instead applies
three residual temporal convolutions of kernel size three with dilations 1, 2 and
4 to the block embedding sequence, then mean pools and reads out through a smal\-ler
stack of its own \cite{bai2018empirical}. Its receptive field of fifteen blocks
covers the longest path we test. Since the whole path is complete at scoring time,
the convolutions are symmetrically padded rather than causal, so path boundaries
enter the ordered network through zero padding as well as through ordering. The
pooled network has 8{,}609 parameters and the ordered network 12{,}929.

The quantity we track is
\begin{equation}
\Delta_{\mathrm{order}} = \widehat I^{\mathrm{ordered}}_{\mathrm{event}}(s_c)
  - \widehat I^{\mathrm{pooled}}_{\mathrm{event}}(s_c),
\end{equation}
the \emph{ordering increment}, reported throughout in clipped nats per event. The two networks are given the same blocks, encoder design, development
split, optimization budget and calibration procedure. The magnitude of the
ordering increment also depends on network size and architecture, which are not
fully matched between the two models.

\subsection{Splits and reporting}
\label{sec:design}

The fifteen days are split chronologically into nine development days, three
calibration days and three held-out test days, the last being 2026-07-21, 22 and
23. Network weights come from development, while the scale $\alpha$ and all
selection thresholds are fixed on calibration. A path, its reverse and every
path that overlaps it stay inside the same split. Optimization uses
AdamW~\cite{loshchilov2019decoupled} with learning rate and weight decay selected
on an expanding-window four-fold split of the development days, for at most thirty
epochs with early stopping. The primary setting of $B=10$ and $K=6$ on NVDA
daytime gives 1{,}751{,}687 development, 671{,}859 calibration and 582{,}767 test
paths.

Overlapping paths are dependent, so a row-level test that assumes independent
samples is inappropriate. Every value below is a held-out mean over the three
test dates weighted equally, and every network is trained three times from
different initializations. A result described as positive throughout is positive
on each test date for each of those initializations, and the twelve-symbol panel
of Section~\ref{sec:cross} provides replication across market conditions.

The primary resolution $B=10$ was fixed before the neural experiments, from a
logistic scan of six block sizes and twelve path lengths. It is the coarsest
setting whose day-level sign stays stable over the full range of $K$ tested,
where $B=2$ becomes unstable beyond about ten updates and $B=30$ and above give a
weak rate per event.

\section{Results}
\label{sec:results}

\subsection{Nonlinearity and temporal ordering}
\label{sec:nonlinear}

We first ask whether a linear antisymmetric score is sufficient to recover the
measured asymmetry. Two antisymmetric logistic scores serve as baselines, one
using the mean, standard deviation, minimum and maximum of each field across the
path and the other the position-wise fields of the flattened path. Even features
take the same values in both orientations, so an order-invariant linear score
drops them entirely. What survives is a fixed weighting of signed returns and net
flows, whose sample means over a session are close to zero. A product of an even
and an odd feature, such as spread times signed net depth change, is odd
under reversal and need not average to zero, but no linear score can form one.

At the primary setting of $B=10$ and $K=6$ for NVDA daytime, the pooled network
reaches 0.0359 nats per event and the ordered one 0.0537. The two logistic
baselines reach 0.000368 and 0.000385, two orders of magnitude below either.
Those two baseline values are unclipped. The measured asymmetry lies almost entirely in nonlinear
composition across path variables. Over one path of about sixty-one realized
events the two network rates are 2.2 and 3.3 nats. The ordering increment is
0.0179, positive throughout and lying between 0.0171 and 0.0187.

\subsection{Where does the increment come from?}
\label{sec:isolating}

Both networks receive the activity clock, but the ordered network can also use
the sequence of its values. Shuffling blocks changes that sequence together
with the order of book states and flows. The ordered network also has more
parameters. Symmetric padding is a third
difference, since only the first and last blocks have zeros in their convolution
window, so the network can locate the path boundary without using order at all.
We take the three in turn.

\paragraph{Block order without the clock.} To test block order without also
scram\-bling e\-lapsed time, we refit both networks
after deleting e\-lapsed time, realized event count and completion
overshoot. In the observed order the remaining 34 features give
0.0126 nats per event for the pooled network and 0.0197 for the ordered one, an
increment of 0.0071. We then randomize the six blocks within each path and refit
the ordered network. It gives 0.0132 against the same pooled value of 0.0126,
which is unchanged by the permutation, leaving an increment of 0.00055. The
reduction shows that most of the clock-free contrast is associated with the
observed block order.

\paragraph{Capacity and boundary.} At $K=1$ a path is a single block and neither
network has any adjacency
available. There the pooled network reaches 0.0232 nats per event and the ordered
one 0.0218, an increment of $-0.00135$ that is negative on every test date and
for every initialization, so the extra capacity of the ordered network does
not by itself produce a positive increment.
Replacing the symmetric padding with a circular variant makes every block position
look interior, and the increment changes only modestly, from 0.0179 to 0.0160.
Together, the shuffle, $K=1$ and padding controls tie the positive contrast to the
observed block order rather than to a generic capacity advantage or a boundary
cue.

\paragraph{Feature groups.} A further question is whether the ordering sits in one
feature group or in their interaction, so we refit both networks on each of the
four groups in turn (Fig.~\ref{fig:identification}(b)). Each group is a separate
refit and the four do not decompose the joint rate. Mechanical flow alone reaches
an ordered rate of 0.0100 and an increment of 0.0053, book state reaches 0.0070
and 0.0029, and liquidity events and price each stay below 0.0010, against a
joint ordered rate of 0.0537 and a joint increment of 0.0179. Removing book state
leaves the ordered rate at 0.0527 while the pooled rate falls from 0.0359 to
0.0224, which raises the increment to 0.0303. No one-group fit reproduces the
full fitted rate. Together with the linear comparisons, these fits show that
the nonlinear models using the joint representation recover more irreversibility
than the linear models or any single feature group. These group refits
remain exploratory, since they followed a screen on the same test dates.

\subsection{Does the increment hold across symbols and sessions?}
\label{sec:cross}

\begin{figure*}[t]
\centering
\includegraphics[width=\textwidth]{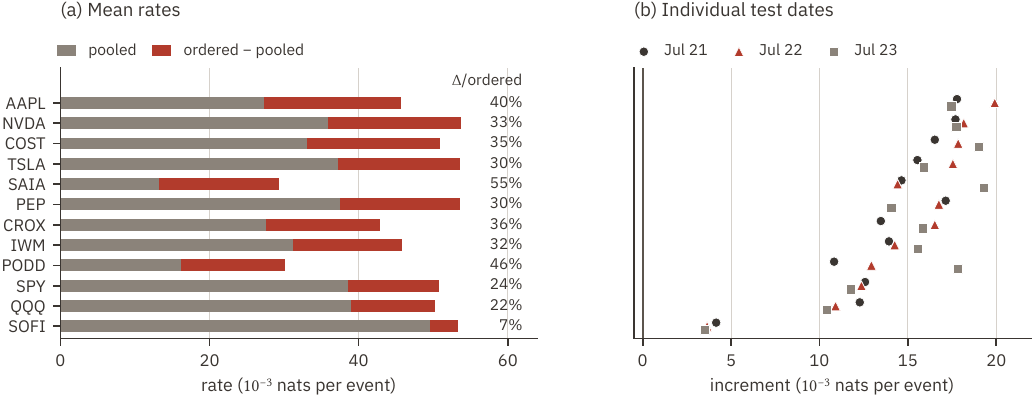}
\Description{Two aligned panels show twelve symbols in decreasing mean ordering increment. At left, stacked bars combine the gray pooled rate and red ordered-minus-pooled increment. Percentages give the increment divided by the ordered rate. At right, a circle, triangle, and square show the increment on July 21, 22, and 23, 2026, each averaged over three network initializations. Every date-specific point is positive. Both horizontal axes use thousandths of a nat per event.}
\caption{Pooled rate and ordering increment for twelve symbols at $B=10$ and
$K=6$ during daytime, in order of decreasing mean increment. In (a), each bar is
the ordered rate split at the pooled value, averaged equally over three test
dates and three initializations. Percentages give
$\Delta_{\mathrm{order}}/\widehat I_{\mathrm{ordered}}$. In (b), each point is
the increment on one test date, averaged over the three initializations. Circles,
triangles and squares denote July 21, 22 and 23, 2026, respectively.}
\label{fig:cross}
\end{figure*}

A property of one symbol during one part of the day would say little about the
book, so for NVDA both networks are refitted and evaluated separately at the
open, before the close and during daytime. The increments are 0.0225, 0.0195 and 0.0187 nats
per event, positive throughout in each window. The last of these comes from a
separately fitted daytime pair and sits close to 0.0179 for the primary fit.
Because the event clock fixes the number of book
updates in a block, these comparisons do not simply reflect different event
counts across the three windows.
Activity, tick constraint and queue depth are the axes along which one book
differs from another, and the twelve symbols span all three, over more than two
orders of magnitude in daily event count. The
mean ordering increment ranges from 0.0038 for SOFI to 0.0184 for AAPL
(Fig.~\ref{fig:cross}(a)). Panel (b) shows the increment on each test date,
averaged over the three initializations. All twelve symbols have positive
increments on each of the three dates.
The rates themselves do not track it, and SOFI reaches one of the highest pooled
rates in the panel while its increment is the smallest. The positive increment
recurs across the twelve selected symbols and the three NVDA session windows.

Within NVDA the pathwise score difference is also not a proxy for realized
variation or event activity. Realized variation alone predicts it with
out-of-sample $R^2=0.0004$, the event rate raises that to 0.0073, and the 29 book
and history inputs of Section~\ref{sec:forecast} reach 0.045 on the same
target.

\subsection{How does the ordering share vary with the spread?}
\label{sec:spread}

\begin{figure}[t]
\centering
\includegraphics[width=\columnwidth]{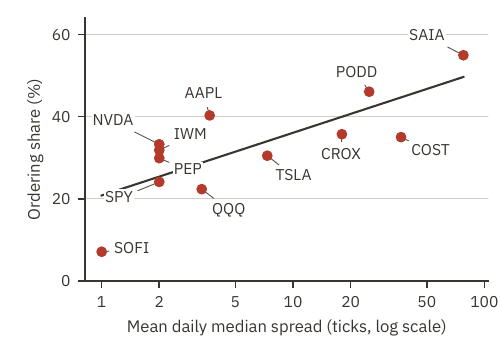}
\Description{A scatterplot shows twelve symbols with spread in ticks on a logarithmic horizontal axis and the ordering increment divided by the ordered rate, in percent, on the vertical axis. Every point is labeled by symbol. The fitted line rises with log spread. SOFI lies near one tick and seven percent, while SAIA lies near seventy-eight ticks and fifty-five percent.}
\caption{Ordering share and spread across the twelve symbols at $B=10$ and
$K=6$ during daytime. Each point uses the same three test dates and three
initializations as Fig.~\ref{fig:cross}. The horizontal coordinate is the
equal-date mean of the daily median block-start spread in ticks. The vertical
coordinate is $100\Delta_{\mathrm{order}}/\widehat I_{\mathrm{ordered}}$,
formed from the date- and initialization-averaged clipped rates. The line is an
unweighted least-squares fit to log spread, with $R^2=0.599$, describing this
cross-section.}
\label{fig:spread}
\end{figure}

A positive increment across symbols does not imply that order contributes
equally in every book. We therefore compare the increment as a fraction of the
ordered rate, which we call the \emph{ordering share},
\begin{equation}
f_{\mathrm{order}}
 = \frac{\Delta_{\mathrm{order}}}{\widehat I_{\mathrm{ordered}}}.
\label{eq:ordering_share}
\end{equation}
The ratio uses the date- and initialization-averaged rates, rather than an
average of pathwise or daily ratios. It compares the fitted networks and is
not an additive decomposition of the true path divergence.

Across the twelve symbols, the ordering share increases roughly linearly with
the logarithm of the spread in ticks (Fig.~\ref{fig:spread}). An unweighted
fit of the percentage share to $\log_{10}$ spread gives a slope of 15.27
percentage points per tenfold difference in spread and $R^2=0.599$. The
association remains positive without clipping, with a slope of 12.50 and
$R^2=0.359$, and the fitted slope stays positive when any one symbol is left
out under either score convention. A larger share need not mean a larger
absolute increment. SOFI has an ordered rate of 0.0533 and a share of 7\%,
whereas SAIA has a lower ordered rate of 0.0293 and a share of 55\%.
This comparison reuses the three evaluation dates of the preceding analyses
and is exploratory.

One interpretation is that the sequence of events matters more when the spread
is wider. A book pinned to one tick is tightly constrained, while a wider spread
may leave more room for different sequences of liquidity changes and price
moves. Under this interpretation, the larger share reflects a greater role for
event order in the observed book dynamics.

Another interpretation concerns estimation. A more varied path distribution may
be harder to learn from a finite sample, and the pooled and ordered networks
need not recover the available information equally well. Their relative
estimation errors can therefore change the share. Separate smaller-sample
refits also changed the share as the number of training paths increased
(Supplementary analysis, Section~S2). Those refits used a different training
protocol, however, so they do not isolate the source of the original
cross-symbol relation. The two interpretations need not be exclusive, and we
have not yet separated these effects. The joint path comparison includes
spread and book dynamics together, without isolating a spread effect.

The shuffled-order refit of Section~\ref{sec:isolating} answers a narrower
question. After removing the clock features, randomizing the six blocks and
retraining reduces the NVDA increment from 0.0071 to 0.00055 nats per event.
The observed block order therefore contributes to the measured contrast.
The test does not change the spread, so it does not yet explain the relation
across symbols. Separating the two interpretations requires matched price
resolution and learning curves under the original training protocol.

\subsection{Is the rate set by the horizon or by the coarse-graining?}
\label{sec:scale}

\begin{figure*}[t]
\centering
\includegraphics[width=\textwidth]{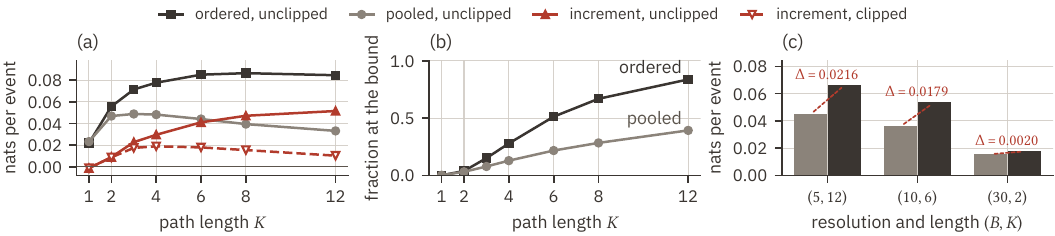}
\Description{Three plots compare path length and event resolution. In panel (a), rates in nats per event are plotted against path length from one to twelve blocks. The ordered rate rises and levels off, the pooled rate peaks early and declines, and the unclipped increment rises while the clipped increment declines at long lengths. Panel (b) shows an increasing fraction of paths at the clipping bound, higher for the ordered readout. Panel (c) compares three resolution and length pairs with the same nominal horizon; dashed connectors mark increments of 0.0216, 0.0179, and 0.0020 from left to right.}
\caption{Dependence of the rate on path length and on event resolution, for NVDA
daytime. In (a), unclipped ordered and pooled rates and their increment are
plotted against path length at $B=10$, with the clipped increment for comparison.
In (b), the fraction of held-out forward paths at the $\log 99$ bound. In (c),
clipped rates at three resolutions holding the nominal horizon at sixty updates,
the dashed connector giving the increment.}
\label{fig:length}
\end{figure*}

The rate is reported per event, so it is fair to ask whether it depends on how
those events are grouped. Path length and the resolution at which blocks are
formed are separate choices, and Fig.~\ref{fig:length} takes them in turn.
For the length comparison, the clipping bound is fixed per path while the
number of events increases with path length. It therefore imposes a
progressively lower ceiling on the rate per event. We examine the unclipped
scores in panel (a) and show the clipped increment alongside them. The fraction
of ordered forward paths at the bound rises from 15\% at $K=3$ to 84\% at
$K=12$, roughly twice the pooled fraction (panel (b)).

The unclipped curves separate as paths lengthen. The ordered rate
reaches 0.0852 nats per event at $K=6$ and stays there, giving 0.0866 at $K=8$
and 0.0847 at $K=12$, while the pooled rate falls from 0.0442 to 0.0395 and then
to 0.0332 over the same range. The ordered rate levels off after roughly forty to
sixty updates and the pooled rate keeps declining, and the primary setting lies at
the short end of that plateau.

An event horizon does not by itself fix the measurement, because the resolution
at which blocks are formed changes it. Panel (c) holds the nominal horizon at 60
updates and uses the clipped rates, as in the preceding comparisons. The pooled
and ordered rates are 0.0447 and 0.0662 at $(B,K)=(5,12)$,
0.0359 and 0.0537 at $(10,6)$, and 0.0156 and 0.0175 at $(30,2)$, so the
increment falls from 0.0216 to 0.0179 and then to 0.0020. At the same length
$K=2$ the increment falls from 0.0087 at $B=10$ to 0.0020 at $B=30$, which
shows that the decrease is not solely a path-length effect. The measured
asymmetry therefore depends on the coarse-graining of the path and not on its
nominal horizon alone. Completion overshoot changes block boundaries, so the
representations at different $B$ are not nested, and the scale profile belongs to
the empirical object itself.

\subsection{Is there irreversibility beyond one step?}
\label{sec:higher}

\begin{figure}[t]
\centering
\includegraphics[width=\columnwidth]{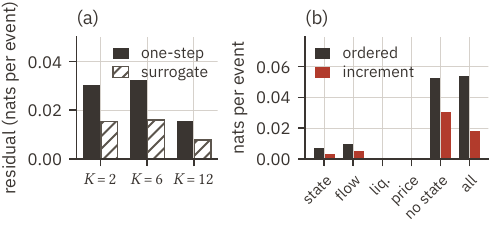}
\Description{Two grouped bar charts use nats per event on their vertical axes. Panel (a) plots the residual above two one-step references at path lengths two, six, and twelve: solid bars for the one-step reference and hatched bars for the surrogate. Both residuals are positive, and the hatched bar is lower at each length. Panel (b) pairs the ordered rate and ordering increment for state, flow, liquidity, price, no-state, and all-feature fits. The no-state and all-feature groups have the tallest ordered-rate bars; their increment bars differ, with the no-state increment larger.}
\caption{One-step references and feature groups at $B=10$. In (a), the observed
long-path score minus each of two one-step references at three path lengths, the
hatched surrogate averaged over twenty randomizations. In (b), the ordered rate
and the ordering increment for each of the four feature groups, namely book
state, mechanical flow, liquidity events and price, for the representation with
book state removed, and for the full representation. Each bar in (b) is a
separate refit, and the bars do not decompose the joint rate.}
\label{fig:identification}
\end{figure}

A first-order process can itself be strongly irreversible whenever $A\to B$ and
$B\to A$ run at different rates, so a positive long-path rate is on its own no
evidence of memory. The rate has to be compared with references that keep only
one-step information. We use two, one restricting what the estimator may see and
the other restricting what the data may contain.

Taking the estimator first, we fit the antisymmetric score at $K=1$, calibrate
it, and sum it along a longer path as $s_{\mathrm{FO}}(W)=\sum_{t}
s_1(S_t,U_{t+1},S_{t+1})$, where $s_1$ is the calibrated $K=1$ score. Summing
that score recovers only part of what the network learns from the whole path. The
shortfall is 0.0305, 0.0324 and 0.0157 nats per event at $K=2$, 6 and 12, on
every test date (Fig.~\ref{fig:identification}(a)).

A shortfall of that kind could still be a limit of the estimator rather than a
property of the book, since a network fitted at $K=1$ never sees a long path and
may simply fail to add up. The surrogate runs the opposite test, leaving the
estimator free while removing the ordering from the data. We map the start and
end of each block to coarse states defined by
clock bin, spread, depth, imbalance, recent activity, recent volatility and
recent duration. Each observed block is then a directed edge between these
states, with bin boundaries fixed from the development days. A randomized
Eulerian traversal reuses every edge exactly once,
preserving state occupancy, directed transition counts and net currents exactly
while changing which transitions follow one another, which is the general
surrogate-data testing principle~\cite{theiler1992testing} applied to the coarse
transition table. If the rate came from that table alone, the surrogate would
reproduce it. Across twenty randomizations the observed
minus surrogate differences are 0.0153, 0.0161 and 0.0078 nats per event at the
same path lengths. The surrogate recovers 0.0376 of the 0.0537 ordered rate at
$K=6$ and leaves 0.0161 nats per event.

The two restrictions are opposite, and both leave a positive residual. The
surrogate residual of 0.0161 falls within a tenth of the 0.0179 ordering
increment obtained from the architecture contrast. Both point at the
same component of the rate, the order in which coarse blocks follow one another,
which neither one-step description retains. We call that component
\emph{higher-order path dependence} in the operational sense of these tests. What
the surrogate residual describes is dependence beyond the one-step transition
table, whether between successive transitions or in continuous path detail lost
by the discrete state map.

% This figure reports the benchmark in Section~\ref{sec:benchmark}.  It is
% declared here so that the full-width float appears beside that discussion
% rather than after the references in the two-column review layout.
\begin{figure*}[t]
\centering
\includegraphics[width=\textwidth]{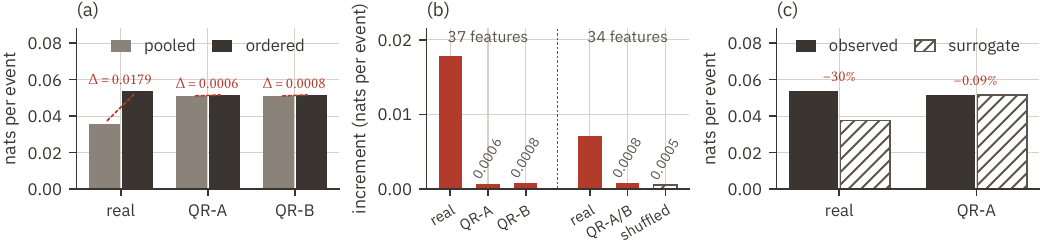}
\Description{Three bar charts compare the real book with queue-reactive simulators. In panel (a), paired pooled and ordered rates nearly coincide for QR-A and QR-B but are separated for the real book; dashed connectors mark the increments. Panel (b) separates the 37-feature comparison from the 34-feature comparison with a vertical dotted line. The real-book increment is smaller with 34 features, while the simulator and shuffled-real bars lie near zero. Panel (c) pairs observed and hatched surrogate rates: the real-book rate falls by 30 percent, whereas the QR-A pair differs by 0.09 percent.}
\caption{Real book and the queue-reactive simulator at $B=10$ and $K=6$. In (a),
pooled and ordered rates for the real book and the two simulator specifications,
the dashed connector giving the ordering increment. In (b), the increment alone,
on all 37 features and on the 34 features excluding the activity clock.
QR-A and QR-B have identical inputs in the latter representation and share one
bar. The hatched bar, labeled shuffled, is the real book refitted with its block
order randomized on the same 34 features. In (c), the coarse
transition surrogate applied to both sources.}
\label{fig:simulator}
\end{figure*}

\subsection{Relation to return prediction}
\label{sec:forecast}

We test whether the pathwise scores add information to standard book and
recent-flow inputs when predicting cumulative midpoint simple returns over the
next 2, 4, 8, 16, 32 and 64 blocks, for NVDA, QQQ, SOFI and SAIA. The
forecaster uses 29 inputs, ten summarizing the current book and nineteen the six most
recent blocks, and is evaluated on walk-forward folds whose training, validation
and refit dates all precede the test date. At the two-block horizon of about
twenty book updates, baseline gradient boosting~\cite{ke2017lightgbm} attains a
median out-of-sample $R^2$ of 0.096 across the four symbols. Adding the ordered
score, the pooled score and their difference changes mean squared error by less
than 0.12\% in either direction in 22 of the 24 symbol--horizon combinations, and
the two larger changes both occur for SAIA on a single test date. Neither these
scores nor alternative normalizations and differences appear to improve on the
standard book and recent-flow inputs, in an exploratory comparison whose
evaluation dates were not held aside from the preceding analyses.

\section{A Benchmark for Order Book Simulators}
\label{sec:benchmark}

The controls above establish that the ordering increment tracks block order in
the real book. A simulator that generates books lets us ask what happens when that
order is produced rather than observed, and this is the gap left open in
Section~\ref{sec:intro}. Existing simulator benchmarks compare generated messages
with real data through dis\-tri\-bu\-tional, impact and dis\-crim\-i\-na\-tor
criteria \cite{nagy2025lobbench},\linebreak[4] and such comparisons do not reach the temporal
asymmetry of the generated sample itself. Comparing each generated path with its
own reverse does.

The queue-reactive simulator is the natural first case, because its event-level
dynamics form a continuous-time Markov chain in the queue state and are therefore
first order by construction, although aggregation over $B$ events can induce
block-level dependence. We calibrate a simplified, fully reinitializing simulator
inspired by Huang et al.~\cite{huang2015simulating} on the same development days,
from queue-size-conditioned limit, cancel and market intensities, their positive
jump-size distributions, and the rate of source events that leave displayed
five-level depth unchanged. The simulated days match the real days exactly in
per-date source-event count. When a best queue empties, the reference price moves
one tick and all ten displayed queues are redrawn independently from the
calibrated limit-arrival size distributions, a reinitialization that is part of
the simulator's relative-price state space.

Two specifications separate content from timing. \mbox{QR-A} keeps the durations
produced by the simulator's own intensities, while \mbox{QR-B} keeps the identical
simulated event content and replaces only its duration sequence with the observed
one, aligned by event ordinal. We refit
and calibrate the networks for each source rather than applying networks trained
on real data to simulated paths.

On the clipped scores at sixty updates, \mbox{QR-A} reaches 0.0510 nats per event
with the pooled network and 0.0517 with the ordered one, against 0.0359 and
0.0537 for the real book. The two sources therefore reach comparable ordered
rates, but the gap between their pooled and ordered estimates is very different
(Fig.~\ref{fig:simulator}(a)): the ordering increment is 0.00065 against
0.0179. \mbox{QR-B} gives 0.00075, so giving simulated content the timing
of the market moves the increment by 0.00010 and does not recover the real-book
value. Deleting the activity clock leaves the same
separation, with a 34-feature increment of 0.0071 for the real book against
0.00078 for both specifications, whose inputs are identical in that
representation. The shuffled-order refit of Section~\ref{sec:isolating} is shown
beside the simulator increments in Fig.~\ref{fig:simulator}(b) as the value the
real book takes when its own ordering is destroyed.

The small simulator increment is not an artifact of features the simulator fails
to produce. Four of the 34 clock-free features are constant in the simulated days
and none is in the real book, since the simulated spread does not move within a day and
best-quote appearance and depletion counts are identically equal. Those four
features lie in three different groups, and in a one-initialization screen
removing any one of those groups from the real representation leaves its
increment at 0.0178 or above. No single missing feature group accounts for the
small simulator increment.

The transition surrogate removes 0.09\% of the simulator's ordered rate against
30\% of the real rate under the same discrete state map
(Fig.~\ref{fig:simulator}(c)). Almost all of the simulated rate is available to
the order-invariant network, and almost none of it depends on which coarse
transition follows which. The calibrated simulator builds the magnitude of the
market's temporal asymmetry out of block composition rather than block order.

The first-order dynamics and queue reinitialization make this simulator a
simple case in which to examine the ordering increment. Learned generators
conditioned on longer event histories could reproduce temporal dependencies
absent from this construction. Whether they recover the real-book ordering
increment remains untested here. Applying the same comparison to their generated
paths would test whether learning from longer histories also reproduces the
additional asymmetry recovered from block order.

\section{Conclusion}

The Nasdaq limit order book paths examined here are statistically irreversible at
event scale. An antisymmetric density-ratio estimator makes this measurable on the
joint path of state and flow, and at six blocks targeting ten source events for
NVDA an order-invariant network reaches 0.036 nats per event while an
order-sensitive one reaches 0.054. The increment of 0.018 keeps its sign for all
twelve symbols on every held-out date, over a panel spanning two orders of
magnitude in activity.

Across the twelve symbols, the increment's share of the ordered rate increases
roughly logarithmically with the spread in ticks. A wider spread may leave more
room for the sequence of events to matter. A more varied path distribution may
also be harder to learn from a finite sample, affecting how much information
each network recovers. We have not yet separated these effects.

Shuffling block order substantially reduces the clock-free increment,
supporting a contribution from the observed ordering.
Randomized paths preserving the coarse one-step transition counts also recover
less irreversibility than the observed paths. The increment changes with the
resolution at which the same event horizon is observed, so comparisons between
books need to specify both their path representation and their observation
scale. The same pathwise scores add no stable forecasting
gain once standard book and recent-flow inputs are present.

A calibrated queue-reactive simulator reaches an ordered rate of 0.052 against
0.054 for the real book at sixty updates, while its ordering increment is
0.0006 against 0.018. Almost all of its measured asymmetry is already recovered
by the order-invariant network. The comparison separates agreement in the
overall rate from agreement in the additional asymmetry recovered by using
block order.

We propose the ordering and scale profile of path irreversibility as a benchmark
for order book simulators. Applied alongside comparisons of generated
and real data, it asks whether a generator reproduces the real book's dependence
on temporal ordering and observation scale.

\section{Disclaimer}
The views expressed are solely those of the authors and do not represent Mirae
Asset Securities. The authors are responsible for any errors or omissions.
This paper is for academic purposes only and does not constitute investment
advice, a recommendation, solicitation, or an offer to buy or sell financial
products or securities.

\begin{acks}
This research was supported by the Basic Science Research Program through the
National Research Foundation of Korea (NRF), funded by the Ministry of Education
(RS-2026-25576625), and by an NRF grant funded by the Korean government (MSIT)
(RS-2023-00242528).
\end{acks}

\bibliographystyle{ACM-Reference-Format}
\bibliography{references}


\begin{thebibliography}{36}

%%% except for \shownote{} and \showURL{}.  The latter does
%%% not use final punctuation, in order to avoid confusing it with
%%% the Web address.

\ifx \showCODEN    \undefined \def \showCODEN     #1{\unskip}     \fi
\ifx \showISBNx    \undefined \def \showISBNx     #1{\unskip}     \fi
\ifx \showISBNxiii \undefined \def \showISBNxiii  #1{\unskip}     \fi
\ifx \showISSN     \undefined \def \showISSN      #1{\unskip}     \fi
\ifx \showLCCN     \undefined \def \showLCCN      #1{\unskip}     \fi
\ifx \shownote     \undefined \def \shownote      #1{#1}          \fi
\ifx \showarticletitle \undefined \def \showarticletitle #1{#1}   \fi
\ifx \showURL      \undefined \def \showURL       {\relax}        \fi
% invisible to TeX
\providecommand\bibfield[2]{#2}
\providecommand\bibinfo[2]{#2}
\providecommand\natexlab[1]{#1}
\providecommand\showeprint[2][]{arXiv:#2}
\makeatletter
\@ifundefined{NAT@parse@date}{}{\let\NAT@parse@date@orig\NAT@parse@date}
\@ifundefined{NAT@parse@date}{}{\def\NAT@parse@date#1#2#3#4#5#6@@{\NAT@parse@date@orig#1#2#3#4#5#6@@\def\NAT@tempyear{0000}\def\NAT@tempexlab{{?}}\ifx\NAT@year\NAT@tempyear\ifx\NAT@exlab\NAT@tempexlab\def\NAT@date{[n.\,d.]}\else\edef\NAT@date{[n.\,d.]\NAT@exlab}\fi\fi}}
\makeatother

\bibitem[An{\'e} and Geman(2000)]%
        {ane2000order}
\bibfield{author}{\bibinfo{person}{Thierry An{\'e}} {and}
  \bibinfo{person}{H{\'e}lyette Geman}.} \bibinfo{year}{2000}\natexlab{}.
\newblock \showarticletitle{Order flow, transaction clock, and normality of
  asset returns}.
\newblock \bibinfo{journal}{\emph{The Journal of Finance}}
  \bibinfo{volume}{55}, \bibinfo{number}{5} (\bibinfo{year}{2000}),
  \bibinfo{pages}{2259--2284}.
\newblock
\href{https://doi.org/10.1111/0022-1082.00286}{doi:\nolinkurl{10.1111/0022-1082.00286}}


\bibitem[Bai et~al\mbox{.}(2018)]%
        {bai2018empirical}
\bibfield{author}{\bibinfo{person}{Shaojie Bai}, \bibinfo{person}{J.~Zico
  Kolter}, {and} \bibinfo{person}{Vladlen Koltun}.}
  \bibinfo{year}{2018}\natexlab{}.
\newblock \bibinfo{title}{An empirical evaluation of generic convolutional and
  recurrent networks for sequence modeling}.
\newblock
\showeprint[arxiv]{1803.01271}~[cs.LG]
\href{https://doi.org/10.48550/arXiv.1803.01271}{doi:\nolinkurl{10.48550/arXiv.1803.01271}}


\bibitem[Bodor and Carlier(2025)]%
        {bodor2025mdqr}
\bibfield{author}{\bibinfo{person}{Hamza Bodor} {and} \bibinfo{person}{Laurent
  Carlier}.} \bibinfo{year}{2025}\natexlab{}.
\newblock \bibinfo{title}{Deep Learning Meets Queue-Reactive: A Framework for
  Realistic Limit Order Book Simulation}.
\newblock
\showeprint[arxiv]{2501.08822}~[q-fin.TR]
\href{https://doi.org/10.48550/arXiv.2501.08822}{doi:\nolinkurl{10.48550/arXiv.2501.08822}}


\bibitem[Bouchaud et~al\mbox{.}(2018)]%
        {bouchaud2018trades}
\bibfield{author}{\bibinfo{person}{Jean-Philippe Bouchaud},
  \bibinfo{person}{Julius Bonart}, \bibinfo{person}{Jonathan Donier}, {and}
  \bibinfo{person}{Martin Gould}.} \bibinfo{year}{2018}\natexlab{}.
\newblock \bibinfo{booktitle}{\emph{Trades, Quotes and Prices: Financial
  Markets Under the Microscope}}.
\newblock \bibinfo{publisher}{Cambridge University Press}.
\newblock
\href{https://doi.org/10.1017/9781316659335}{doi:\nolinkurl{10.1017/9781316659335}}


\bibitem[Cont et~al\mbox{.}(2023)]%
        {cont2023lobgan}
\bibfield{author}{\bibinfo{person}{Rama Cont}, \bibinfo{person}{Mihai
  Cucuringu}, \bibinfo{person}{Jonathan Kochems}, {and} \bibinfo{person}{Felix
  Prenzel}.} \bibinfo{year}{2023}\natexlab{}.
\newblock \bibinfo{title}{Limit Order Book Simulation with Generative
  Adversarial Networks}.
\newblock \bibinfo{howpublished}{SSRN preprint 4512356}.
\newblock
\href{https://doi.org/10.2139/ssrn.4512356}{doi:\nolinkurl{10.2139/ssrn.4512356}}


\bibitem[Cont and de~Larrard(2013)]%
        {cont2013price}
\bibfield{author}{\bibinfo{person}{Rama Cont} {and} \bibinfo{person}{Adrien de
  Larrard}.} \bibinfo{year}{2013}\natexlab{}.
\newblock \showarticletitle{Price dynamics in a {M}arkovian limit order
  market}.
\newblock \bibinfo{journal}{\emph{SIAM Journal on Financial Mathematics}}
  \bibinfo{volume}{4}, \bibinfo{number}{1} (\bibinfo{year}{2013}),
  \bibinfo{pages}{1--25}.
\newblock
\href{https://doi.org/10.1137/110856605}{doi:\nolinkurl{10.1137/110856605}}


\bibitem[Cont et~al\mbox{.}(2014)]%
        {cont2014price}
\bibfield{author}{\bibinfo{person}{Rama Cont}, \bibinfo{person}{Arseniy
  Kukanov}, {and} \bibinfo{person}{Sasha Stoikov}.}
  \bibinfo{year}{2014}\natexlab{}.
\newblock \showarticletitle{The price impact of order book events}.
\newblock \bibinfo{journal}{\emph{Journal of Financial Econometrics}}
  \bibinfo{volume}{12}, \bibinfo{number}{1} (\bibinfo{year}{2014}),
  \bibinfo{pages}{47--88}.
\newblock
\href{https://doi.org/10.1093/jjfinec/nbt003}{doi:\nolinkurl{10.1093/jjfinec/nbt003}}


\bibitem[Cordi et~al\mbox{.}(2021)]%
        {cordi2021nanostructure}
\bibfield{author}{\bibinfo{person}{Marcus Cordi}, \bibinfo{person}{Damien
  Challet}, {and} \bibinfo{person}{Serge Kassibrakis}.}
  \bibinfo{year}{2021}\natexlab{}.
\newblock \showarticletitle{The market nanostructure origin of asset price time
  reversal asymmetry}.
\newblock \bibinfo{journal}{\emph{Quantitative Finance}} \bibinfo{volume}{21},
  \bibinfo{number}{2} (\bibinfo{year}{2021}), \bibinfo{pages}{295--304}.
\newblock
\href{https://doi.org/10.1080/14697688.2020.1753883}{doi:\nolinkurl{10.1080/14697688.2020.1753883}}


\bibitem[Cordi et~al\mbox{.}(2018)]%
        {cordi2017testing}
\bibfield{author}{\bibinfo{person}{Marcus Cordi}, \bibinfo{person}{Damien
  Challet}, {and} \bibinfo{person}{Ioane Muni~Toke}.}
  \bibinfo{year}{2018}\natexlab{}.
\newblock \showarticletitle{Testing the causality of {H}awkes processes with
  time reversal}.
\newblock \bibinfo{journal}{\emph{Journal of Statistical Mechanics: Theory and
  Experiment}} \bibinfo{volume}{2018}, \bibinfo{number}{3}
  (\bibinfo{year}{2018}), \bibinfo{pages}{033408}.
\newblock
\href{https://doi.org/10.1088/1742-5468/aaac3f}{doi:\nolinkurl{10.1088/1742-5468/aaac3f}}


\bibitem[Eisler et~al\mbox{.}(2012)]%
        {eisler2012price}
\bibfield{author}{\bibinfo{person}{Zolt{\'a}n Eisler},
  \bibinfo{person}{Jean-Philippe Bouchaud}, {and} \bibinfo{person}{Julien
  Kockelkoren}.} \bibinfo{year}{2012}\natexlab{}.
\newblock \showarticletitle{The price impact of order book events: market
  orders, limit orders and cancellations}.
\newblock \bibinfo{journal}{\emph{Quantitative Finance}} \bibinfo{volume}{12},
  \bibinfo{number}{9} (\bibinfo{year}{2012}), \bibinfo{pages}{1395--1419}.
\newblock
\href{https://doi.org/10.1080/14697688.2010.528444}{doi:\nolinkurl{10.1080/14697688.2010.528444}}


\bibitem[Esposito(2012)]%
        {esposito2012stochastic}
\bibfield{author}{\bibinfo{person}{Massimiliano Esposito}.}
  \bibinfo{year}{2012}\natexlab{}.
\newblock \showarticletitle{Stochastic thermodynamics under coarse graining}.
\newblock \bibinfo{journal}{\emph{Physical Review E}} \bibinfo{volume}{85},
  \bibinfo{number}{4} (\bibinfo{year}{2012}), \bibinfo{pages}{041125}.
\newblock
\href{https://doi.org/10.1103/PhysRevE.85.041125}{doi:\nolinkurl{10.1103/PhysRevE.85.041125}}


\bibitem[Flanagan and Lacasa(2016)]%
        {flanagan2016irreversibility}
\bibfield{author}{\bibinfo{person}{Ryan Flanagan} {and} \bibinfo{person}{Lucas
  Lacasa}.} \bibinfo{year}{2016}\natexlab{}.
\newblock \showarticletitle{Irreversibility of financial time series: A
  graph-theoretical approach}.
\newblock \bibinfo{journal}{\emph{Physics Letters A}} \bibinfo{volume}{380},
  \bibinfo{number}{20} (\bibinfo{year}{2016}), \bibinfo{pages}{1689--1697}.
\newblock
\href{https://doi.org/10.1016/j.physleta.2016.03.011}{doi:\nolinkurl{10.1016/j.physleta.2016.03.011}}


\bibitem[Frey et~al\mbox{.}(2023)]%
        {frey2023jaxlob}
\bibfield{author}{\bibinfo{person}{Sascha~Yves Frey}, \bibinfo{person}{Kang
  Li}, \bibinfo{person}{Peer Nagy}, \bibinfo{person}{Silvia Sapora},
  \bibinfo{person}{Christopher Lu}, \bibinfo{person}{Stefan Zohren},
  \bibinfo{person}{Jakob~Nicolaus Foerster}, {and} \bibinfo{person}{Anisoara
  Calinescu}.} \bibinfo{year}{2023}\natexlab{}.
\newblock \showarticletitle{{JAX-LOB}: A {GPU}-Accelerated Limit Order Book
  Simulator to Unlock Large Scale Reinforcement Learning for Trading}. In
  \bibinfo{booktitle}{\emph{Proceedings of the Fourth ACM International
  Conference on AI in Finance}}. \bibinfo{publisher}{ACM},
  \bibinfo{pages}{583--591}.
\newblock
\href{https://doi.org/10.1145/3604237.3626880}{doi:\nolinkurl{10.1145/3604237.3626880}}


\bibitem[Gould et~al\mbox{.}(2013)]%
        {gould2013limit}
\bibfield{author}{\bibinfo{person}{Martin~D. Gould}, \bibinfo{person}{Mason~A.
  Porter}, \bibinfo{person}{Stacy Williams}, \bibinfo{person}{Mark McDonald},
  \bibinfo{person}{Daniel~J. Fenn}, {and} \bibinfo{person}{Sam~D. Howison}.}
  \bibinfo{year}{2013}\natexlab{}.
\newblock \showarticletitle{Limit order books}.
\newblock \bibinfo{journal}{\emph{Quantitative Finance}} \bibinfo{volume}{13},
  \bibinfo{number}{11} (\bibinfo{year}{2013}), \bibinfo{pages}{1709--1742}.
\newblock
\href{https://doi.org/10.1080/14697688.2013.803148}{doi:\nolinkurl{10.1080/14697688.2013.803148}}


\bibitem[Guo et~al\mbox{.}(2017)]%
        {guo2017calibration}
\bibfield{author}{\bibinfo{person}{Chuan Guo}, \bibinfo{person}{Geoff Pleiss},
  \bibinfo{person}{Yu Sun}, {and} \bibinfo{person}{Kilian~Q. Weinberger}.}
  \bibinfo{year}{2017}\natexlab{}.
\newblock \showarticletitle{On calibration of modern neural networks}. In
  \bibinfo{booktitle}{\emph{Proceedings of the 34th International Conference on
  Machine Learning}} \emph{(\bibinfo{series}{Proceedings of Machine Learning
  Research}, Vol.~\bibinfo{volume}{70})}. \bibinfo{publisher}{PMLR},
  \bibinfo{pages}{1321--1330}.
\newblock
\urldef\tempurl%
\url{https://proceedings.mlr.press/v70/guo17a.html}
\showURL{%
\tempurl}


\bibitem[Gutmann and Hyv{\"a}rinen(2010)]%
        {gutmann2010noise}
\bibfield{author}{\bibinfo{person}{Michael Gutmann} {and} \bibinfo{person}{Aapo
  Hyv{\"a}rinen}.} \bibinfo{year}{2010}\natexlab{}.
\newblock \showarticletitle{Noise-contrastive estimation: A new estimation
  principle for unnormalized statistical models}. In
  \bibinfo{booktitle}{\emph{Proceedings of the 13th International Conference on
  Artificial Intelligence and Statistics}} \emph{(\bibinfo{series}{Proceedings
  of Machine Learning Research}, Vol.~\bibinfo{volume}{9})}.
  \bibinfo{publisher}{PMLR}, \bibinfo{address}{Chia Laguna Resort, Sardinia,
  Italy}, \bibinfo{pages}{297--304}.
\newblock
\urldef\tempurl%
\url{https://proceedings.mlr.press/v9/gutmann10a.html}
\showURL{%
\tempurl}


\bibitem[Huang et~al\mbox{.}(2015)]%
        {huang2015simulating}
\bibfield{author}{\bibinfo{person}{Weibing Huang},
  \bibinfo{person}{Charles-Albert Lehalle}, {and} \bibinfo{person}{Mathieu
  Rosenbaum}.} \bibinfo{year}{2015}\natexlab{}.
\newblock \showarticletitle{Simulating and analyzing order book data: The
  queue-reactive model}.
\newblock \bibinfo{journal}{\emph{J. Amer. Statist. Assoc.}}
  \bibinfo{volume}{110}, \bibinfo{number}{509} (\bibinfo{year}{2015}),
  \bibinfo{pages}{107--122}.
\newblock
\href{https://doi.org/10.1080/01621459.2014.982278}{doi:\nolinkurl{10.1080/01621459.2014.982278}}


\bibitem[Ke et~al\mbox{.}(2017)]%
        {ke2017lightgbm}
\bibfield{author}{\bibinfo{person}{Guolin Ke}, \bibinfo{person}{Qi Meng},
  \bibinfo{person}{Thomas Finley}, \bibinfo{person}{Taifeng Wang},
  \bibinfo{person}{Wei Chen}, \bibinfo{person}{Weidong Ma},
  \bibinfo{person}{Qiwei Ye}, {and} \bibinfo{person}{Tie-Yan Liu}.}
  \bibinfo{year}{2017}\natexlab{}.
\newblock \showarticletitle{{LightGBM}: A highly efficient gradient boosting
  decision tree}. In \bibinfo{booktitle}{\emph{Advances in Neural Information
  Processing Systems}}, Vol.~\bibinfo{volume}{30}. \bibinfo{publisher}{Curran
  Associates, Inc.}
\newblock
\urldef\tempurl%
\url{https://proceedings.neurips.cc/paper_files/paper/2017/hash/6449f44a102fde848669bdd9eb6b76fa-Abstract.html}
\showURL{%
\tempurl}


\bibitem[Kim et~al\mbox{.}(2020)]%
        {kim2020learning}
\bibfield{author}{\bibinfo{person}{Dong-Kyum Kim}, \bibinfo{person}{Youngkyoung
  Bae}, \bibinfo{person}{Sangyun Lee}, {and} \bibinfo{person}{Hawoong Jeong}.}
  \bibinfo{year}{2020}\natexlab{}.
\newblock \showarticletitle{Learning entropy production via neural networks}.
\newblock \bibinfo{journal}{\emph{Physical Review Letters}}
  \bibinfo{volume}{125}, \bibinfo{number}{14} (\bibinfo{year}{2020}),
  \bibinfo{pages}{140604}.
\newblock
\href{https://doi.org/10.1103/PhysRevLett.125.140604}{doi:\nolinkurl{10.1103/PhysRevLett.125.140604}}


\bibitem[Lacasa et~al\mbox{.}(2012)]%
        {lacasa2012time}
\bibfield{author}{\bibinfo{person}{Lucas Lacasa}, \bibinfo{person}{Angel
  Nu{\~n}ez}, \bibinfo{person}{{\'E}dgar Rold{\'a}n}, \bibinfo{person}{Juan
  M.~R. Parrondo}, {and} \bibinfo{person}{Bartolo Luque}.}
  \bibinfo{year}{2012}\natexlab{}.
\newblock \showarticletitle{Time series irreversibility: a visibility graph
  approach}.
\newblock \bibinfo{journal}{\emph{The European Physical Journal B}}
  \bibinfo{volume}{85}, \bibinfo{number}{6} (\bibinfo{year}{2012}),
  \bibinfo{pages}{217}.
\newblock
\href{https://doi.org/10.1140/epjb/e2012-20809-8}{doi:\nolinkurl{10.1140/epjb/e2012-20809-8}}


\bibitem[Loshchilov and Hutter(2019)]%
        {loshchilov2019decoupled}
\bibfield{author}{\bibinfo{person}{Ilya Loshchilov} {and}
  \bibinfo{person}{Frank Hutter}.} \bibinfo{year}{2019}\natexlab{}.
\newblock \showarticletitle{Decoupled weight decay regularization}. In
  \bibinfo{booktitle}{\emph{International Conference on Learning
  Representations}}.
\newblock
\urldef\tempurl%
\url{https://openreview.net/forum?id=Bkg6RiCqY7}
\showURL{%
\tempurl}


\bibitem[Nagy et~al\mbox{.}(2025)]%
        {nagy2025lobbench}
\bibfield{author}{\bibinfo{person}{Peer Nagy}, \bibinfo{person}{Sascha~Yves
  Frey}, \bibinfo{person}{Kang Li}, \bibinfo{person}{Bidipta Sarkar},
  \bibinfo{person}{Svitlana Vyetrenko}, \bibinfo{person}{Stefan Zohren},
  \bibinfo{person}{Ani Calinescu}, {and} \bibinfo{person}{Jakob~Nicolaus
  Foerster}.} \bibinfo{year}{2025}\natexlab{}.
\newblock \showarticletitle{{LOB-Bench}: Benchmarking Generative {AI} for
  Finance, an Application to Limit Order Book Data}. In
  \bibinfo{booktitle}{\emph{Proceedings of the 42nd International Conference on
  Machine Learning}} \emph{(\bibinfo{series}{Proceedings of Machine Learning
  Research}, Vol.~\bibinfo{volume}{267})}. \bibinfo{publisher}{PMLR},
  \bibinfo{pages}{45437--45460}.
\newblock


\bibitem[Noble et~al\mbox{.}(2026)]%
        {noble2026reality}
\bibfield{author}{\bibinfo{person}{Patrick Noble}, \bibinfo{person}{Mathieu
  Rosenbaum}, {and} \bibinfo{person}{Saad Souilmi}.}
  \bibinfo{year}{2026}\natexlab{}.
\newblock \bibinfo{title}{Bridging the Reality Gap in Limit Order Book
  Simulation}.
\newblock
\showeprint[arxiv]{2603.24137}~[q-fin.TR]
\href{https://doi.org/10.48550/arXiv.2603.24137}{doi:\nolinkurl{10.48550/arXiv.2603.24137}}


\bibitem[Ohga et~al\mbox{.}(2023)]%
        {ohga2023thermodynamic}
\bibfield{author}{\bibinfo{person}{Naruo Ohga}, \bibinfo{person}{Sosuke Ito},
  {and} \bibinfo{person}{Artemy Kolchinsky}.} \bibinfo{year}{2023}\natexlab{}.
\newblock \showarticletitle{Thermodynamic bound on the asymmetry of
  cross-correlations}.
\newblock \bibinfo{journal}{\emph{Physical Review Letters}}
  \bibinfo{volume}{131}, \bibinfo{number}{7} (\bibinfo{year}{2023}),
  \bibinfo{pages}{077101}.
\newblock
\href{https://doi.org/10.1103/PhysRevLett.131.077101}{doi:\nolinkurl{10.1103/PhysRevLett.131.077101}}


\bibitem[Parrondo et~al\mbox{.}(2009)]%
        {parrondo2009entropy}
\bibfield{author}{\bibinfo{person}{Juan M.~R. Parrondo},
  \bibinfo{person}{Christian Van~den Broeck}, {and} \bibinfo{person}{Ryoichi
  Kawai}.} \bibinfo{year}{2009}\natexlab{}.
\newblock \showarticletitle{Entropy production and the arrow of time}.
\newblock \bibinfo{journal}{\emph{New Journal of Physics}}
  \bibinfo{volume}{11}, \bibinfo{number}{7} (\bibinfo{year}{2009}),
  \bibinfo{pages}{073008}.
\newblock
\href{https://doi.org/10.1088/1367-2630/11/7/073008}{doi:\nolinkurl{10.1088/1367-2630/11/7/073008}}


\bibitem[Rold{\'a}n and Parrondo(2010)]%
        {roldan2010estimating}
\bibfield{author}{\bibinfo{person}{{\'E}dgar Rold{\'a}n} {and}
  \bibinfo{person}{Juan M.~R. Parrondo}.} \bibinfo{year}{2010}\natexlab{}.
\newblock \showarticletitle{Estimating dissipation from single stationary
  trajectories}.
\newblock \bibinfo{journal}{\emph{Physical Review Letters}}
  \bibinfo{volume}{105}, \bibinfo{number}{15} (\bibinfo{year}{2010}),
  \bibinfo{pages}{150607}.
\newblock
\href{https://doi.org/10.1103/PhysRevLett.105.150607}{doi:\nolinkurl{10.1103/PhysRevLett.105.150607}}


\bibitem[Seifert(2012)]%
        {seifert2012stochastic}
\bibfield{author}{\bibinfo{person}{Udo Seifert}.}
  \bibinfo{year}{2012}\natexlab{}.
\newblock \showarticletitle{Stochastic thermodynamics, fluctuation theorems and
  molecular machines}.
\newblock \bibinfo{journal}{\emph{Reports on Progress in Physics}}
  \bibinfo{volume}{75}, \bibinfo{number}{12} (\bibinfo{year}{2012}),
  \bibinfo{pages}{126001}.
\newblock
\href{https://doi.org/10.1088/0034-4885/75/12/126001}{doi:\nolinkurl{10.1088/0034-4885/75/12/126001}}


\bibitem[Sitaru et~al\mbox{.}(2023)]%
        {sitaru2023orderflow}
\bibfield{author}{\bibinfo{person}{Bogdan Sitaru}, \bibinfo{person}{Anisoara
  Calinescu}, {and} \bibinfo{person}{Mihai Cucuringu}.}
  \bibinfo{year}{2023}\natexlab{}.
\newblock \showarticletitle{Order Flow Decomposition for Price Impact Analysis
  in Equity Limit Order Books}. In \bibinfo{booktitle}{\emph{Proceedings of the
  Fourth ACM International Conference on AI in Finance}}.
  \bibinfo{publisher}{ACM}, \bibinfo{pages}{637--645}.
\newblock
\href{https://doi.org/10.1145/3604237.3626874}{doi:\nolinkurl{10.1145/3604237.3626874}}


\bibitem[Sugiyama et~al\mbox{.}(2012)]%
        {sugiyama2012density}
\bibfield{author}{\bibinfo{person}{Masashi Sugiyama}, \bibinfo{person}{Taiji
  Suzuki}, {and} \bibinfo{person}{Takafumi Kanamori}.}
  \bibinfo{year}{2012}\natexlab{}.
\newblock \bibinfo{booktitle}{\emph{Density Ratio Estimation in Machine
  Learning}}.
\newblock \bibinfo{publisher}{Cambridge University Press}.
\newblock
\href{https://doi.org/10.1017/CBO9781139035613}{doi:\nolinkurl{10.1017/CBO9781139035613}}


\bibitem[Theiler et~al\mbox{.}(1992)]%
        {theiler1992testing}
\bibfield{author}{\bibinfo{person}{James Theiler}, \bibinfo{person}{Stephen
  Eubank}, \bibinfo{person}{Andr{\'e} Longtin}, \bibinfo{person}{Bryan
  Galdrikian}, {and} \bibinfo{person}{J.~Doyne Farmer}.}
  \bibinfo{year}{1992}\natexlab{}.
\newblock \showarticletitle{Testing for nonlinearity in time series: the method
  of surrogate data}.
\newblock \bibinfo{journal}{\emph{Physica D: Nonlinear Phenomena}}
  \bibinfo{volume}{58}, \bibinfo{number}{1-4} (\bibinfo{year}{1992}),
  \bibinfo{pages}{77--94}.
\newblock
\href{https://doi.org/10.1016/0167-2789(92)90102-S}{doi:\nolinkurl{10.1016/0167-2789(92)90102-S}}


\bibitem[Vodret et~al\mbox{.}(2024)]%
        {vodret2024functional}
\bibfield{author}{\bibinfo{person}{Michele Vodret}, \bibinfo{person}{Cristiano
  Pacini}, {and} \bibinfo{person}{Christian Bongiorno}.}
  \bibinfo{year}{2024}\natexlab{}.
\newblock \showarticletitle{Functional decomposition and estimation of
  irreversibility in time series via machine learning}.
\newblock \bibinfo{journal}{\emph{Physical Review E}} \bibinfo{volume}{110},
  \bibinfo{number}{6} (\bibinfo{year}{2024}), \bibinfo{pages}{064310}.
\newblock
\href{https://doi.org/10.1103/PhysRevE.110.064310}{doi:\nolinkurl{10.1103/PhysRevE.110.064310}}


\bibitem[Weiss(1975)]%
        {weiss1975time}
\bibfield{author}{\bibinfo{person}{Gideon Weiss}.}
  \bibinfo{year}{1975}\natexlab{}.
\newblock \showarticletitle{Time-reversibility of linear stochastic processes}.
\newblock \bibinfo{journal}{\emph{Journal of Applied Probability}}
  \bibinfo{volume}{12}, \bibinfo{number}{4} (\bibinfo{year}{1975}),
  \bibinfo{pages}{831--836}.
\newblock
\href{https://doi.org/10.2307/3212735}{doi:\nolinkurl{10.2307/3212735}}


\bibitem[Zaheer et~al\mbox{.}(2017)]%
        {zaheer2017deep}
\bibfield{author}{\bibinfo{person}{Manzil Zaheer}, \bibinfo{person}{Satwik
  Kottur}, \bibinfo{person}{Siamak Ravanbakhsh}, \bibinfo{person}{Barnabas
  Poczos}, \bibinfo{person}{Ruslan Salakhutdinov}, {and}
  \bibinfo{person}{Alexander Smola}.} \bibinfo{year}{2017}\natexlab{}.
\newblock \showarticletitle{Deep sets}. In \bibinfo{booktitle}{\emph{Advances
  in Neural Information Processing Systems}}, Vol.~\bibinfo{volume}{30}.
  \bibinfo{publisher}{Curran Associates, Inc.}
\newblock
\urldef\tempurl%
\url{https://proceedings.neurips.cc/paper_files/paper/2017/hash/f22e4747da1aa27e363d86d40ff442fe-Abstract.html}
\showURL{%
\tempurl}


\bibitem[Zhang et~al\mbox{.}(2026)]%
        {zhang2025clusterlob}
\bibfield{author}{\bibinfo{person}{Yichi Zhang}, \bibinfo{person}{Mihai
  Cucuringu}, \bibinfo{person}{Alexander~Y. Shestopaloff}, {and}
  \bibinfo{person}{Stefan Zohren}.} \bibinfo{year}{2026}\natexlab{}.
\newblock \showarticletitle{{ClusterLOB}: enhancing trading strategies by
  clustering orders in limit order books}.
\newblock \bibinfo{journal}{\emph{Quantitative Finance}} \bibinfo{volume}{26},
  \bibinfo{number}{8} (\bibinfo{year}{2026}), \bibinfo{pages}{1285--1309}.
\newblock
\href{https://doi.org/10.1080/14697688.2026.2665153}{doi:\nolinkurl{10.1080/14697688.2026.2665153}}


\bibitem[Zhang et~al\mbox{.}(2019)]%
        {zhang2019deeplob}
\bibfield{author}{\bibinfo{person}{Zihao Zhang}, \bibinfo{person}{Stefan
  Zohren}, {and} \bibinfo{person}{Stephen Roberts}.}
  \bibinfo{year}{2019}\natexlab{}.
\newblock \showarticletitle{{DeepLOB}: Deep convolutional neural networks for
  limit order books}.
\newblock \bibinfo{journal}{\emph{IEEE Transactions on Signal Processing}}
  \bibinfo{volume}{67}, \bibinfo{number}{11} (\bibinfo{year}{2019}),
  \bibinfo{pages}{3001--3012}.
\newblock
\href{https://doi.org/10.1109/TSP.2019.2907260}{doi:\nolinkurl{10.1109/TSP.2019.2907260}}


\bibitem[Zumbach(2009)]%
        {zumbach2009time}
\bibfield{author}{\bibinfo{person}{Gilles Zumbach}.}
  \bibinfo{year}{2009}\natexlab{}.
\newblock \showarticletitle{Time reversal invariance in finance}.
\newblock \bibinfo{journal}{\emph{Quantitative Finance}} \bibinfo{volume}{9},
  \bibinfo{number}{5} (\bibinfo{year}{2009}), \bibinfo{pages}{505--515}.
\newblock
\href{https://doi.org/10.1080/14697680802616712}{doi:\nolinkurl{10.1080/14697680802616712}}


\end{thebibliography}

\end{document}